\documentclass[twocolumn,twoside,slac]{revtex4}
\usepackage{graphicx}
\usepackage{fancyhdr}
\renewcommand{\headrulewidth}{0pt}
\renewcommand{\footrulewidth}{0pt}

\begin{document}

\title{Concrete uses of XML in software
development and data analysis.}

%

\author{S. Patton}
\affiliation{LBNL, Berkeley, CA 94720, USA}

\begin{abstract}
XML is now becoming an industry standard for data description and
exchange. Despite this there are still some questions about how or if
this technology can be useful in High Energy Physics software
development and data analysis. This paper aims to answer these questions
by demonstrating how XML is used in the IceCube software development
system, data handling and analysis. It does this by first surveying
the concepts and tools that make up the XML technology. It then goes on
to discuss concrete examples of how these concepts and tools are used to
speed up software development in IceCube and what are the benefits of
using XML in IceCube's data handling and analysis chain. The overall aim
of this paper it to show that XML does have many benefits to bring High
Energy Physics software development and data analysis.
\end{abstract}

\maketitle

\thispagestyle{fancy}


\section{Introduction}

Every few years there is a new technology that is touted as solving all
our computing problems. In some cases these technologies turn out to be
much less than they appeared to be, while other technologies do
substantially advance our profession. XML is currently one of these
technologies {\em du-jour}. As with all other technologies XML's
usefulness can only be determined once there has been some concrete
implementations of the technology. This aim of this paper is to report
concrete uses of XML within the IceCube experiment, and to demonstrate
what benefits this technology actually delivers.

\subsection{IceCube Overview}

The concrete examples of XML given in this paper are taken from the
IceCube experiment. This experiment is a Neutrino telescope that will be
deployed at the South Pole. It is the successor to the Amanda experiment
which has been taking data since the mid-to-late 90's. The experiment
itself is relatively small in H.E.P. terms today  with the collaboration
comprising of around 150 physicists. This small size has enhanced the
need for it to look to others for its software needs rather than to
develop all its own software from the ground up. As will be seen
throughout the rest of this paper the use XML has helped satisfy this
requirement.

IceCube's adoption of XML has also been facilitated by two other issues.
The first is that there is very little legacy code or data from the
Amanda experiment. What code there is is mainly restricted to the
simulation portion of the software and any legacy data is in a home
grown ASCII format which can be easy converted into XML. The second
issue was the decision IceCube made that the core DAQ and data handling
software should be written in Java. It turns out that many of the tools
for, and advancements in, XML are taking place in Java first and then
being migrated to other languages. This places IceCube is a good
position to exploit the latest XML technologies.

\section{XML}

\subsection{Overview}

XML (eXtensible Markup Language) is a derivative of SGML (Standard
Generalized Markup Language). While SGML is, as it name suggests, very
general this generality also means that it can be very time consuming to
use it to implement solutions. At the other end of the spectrum a markup
language such as HTML (HyperText Markup Language) is completely defined
and while it is easy to use it is impossible to adapt to different
needs. XML aims to take a middle road between the extremes of SGML and
HTML by providing a flexible, customizable markup language that is easy
to use.

The core idea of XML is that is it a text base language that defines the
grammar of documents. This allows for the separation of the structure of
a document from its content. Being text based also means that it can be
independent of programming languages making it an ideal candidate for
data transfer between programs. In fact this has lead to it becoming an
industry standard for data description and exchange.

\subsection{Standards}

One of the reasons XML has been successful at becoming an industry
standard is that many aspects of its technology have become standards
themselves, either as W3C~\cite{w3c-ref} recommendations or as {\em
de-facto} standards due to widespread adoption. The following is a list
of the various area of XML technology and the standards which apply to
them (The {\em emphasized} standards are W3C standards.)

\begin{figure*}[t]
\centering
\begin{verbatim}
<?xml version="1.0" encoding="UTF-8"?>
<daq:AtwdReadout
        xmlns:daq = "http://glacier.lbl.gov/icecube/daq/example"
        xmlns:xsi = "http://www.w3.org/2001/XMLSchema-instance"
        xsi:schemaLocation = "http://glacier.lbl.gov/icecube/daq/example atwdReadout.xsd">
  <Atwd>
    <Channel number="0" bitsPerSample="8">
      67 71 72 68 73 69 71 70 77 71 72 70
      73 75 78 76 77 80 71 73 82 75 79 78
      76 81 75 85 81 86 82 86 81 84 79 80
      84 188 0 0 0 0 0 0 0 0 0 0
    </Channel>
    <Channel number="2">
      231 1010 1021 253 995 1021 1021 1021 1021 253 987 1021
      253 221 1000 253 229 982 253 219 211 1009 1021 1021
      1021 253 985 1021 1021 1021 1021 1021 1021 1021 1021 1021
      253 28 82 79 71 71 77 71 77 71 73 75
    </Channel>
   </Atwd>
</daq:AtwdReadout>
\end{verbatim}
\caption{Example XML file {\tt atwdExample.xml}.}
\label{atwdExample-xml}
\end{figure*}

\begin{description}
\item[Parsing] {\em Document Object Model (DOM)} and Simple API for XML
(SAX).
\item[Validation] {\em XML Schema} and Document Type Definitions (DTD).
\item[Language Data Structures] Java Architecture for XML Binding (JAXB).
\item[Transforms] {\em eXtensible Stylesheet Language (XSL)} and {\em 
XSL Transformations (XSLT)}.
\item[Searching] {\em XPath} and {\em XQuery}.
\item[Data Transfers] {\em SOAP}.
\end{description}

Many of these areas will be covered in the rest of this paper.

\section{Parsing}

\subsection{Initial Document Creation}

The aim of parsing a document is to read its contents and make it
available to a program. To start this process it is necessary to have
the XML document in the first place. Once a system is up and running the
creation of such documents is normally straight forward, but during the
development phase of a system the documents are often hand-crafted. The
approach taken on IceCube is that we start of with a complete prose
description of the contents we need in a document. We then create an XML
document which contains this data and is structured into elements which
map onto data objects discussed in the description.
Figure~\ref{atwdExample-xml} is a simplified example of such a document.
This document shows the readout of a two channel ATWD (Analog Transient
Waveform Digitizer).

\subsection{Document Object Model}

There are two standard approaches to parsing an XML document. The first
is the Document Object Model (DOM), which is a W3C standard. In this
approach the whole document is parsed all at once and a representation,
using the ``Composite''~\cite{gamma-ref} design pattern, is created in
memory. It is then up to the program to navigate around this memory
structure to access the information that it needs. While this approach
is very simple from the developers point of view, custom representation
of the data require a second round of processing, accessing the original
memory structure, to transform a document's content in to anything other
than a composite tree. Moreover for large documents this approach can
end up being a resource hog as the whole document appears in memory.

\subsection{Simple API for XML}

The alternate approach to parsing is the Simple API for XML (SAX). In
this approach each element is handed to a call-back routine for
processing. This requires that the developer of these handlers to
maintain the document context as it is processed. On the other hand this
allows for the processing of content as it is read and so the full
document may not be held in memory after parsing has completed.

\subsection{Tools}

There are a number of different tools available that can handle the
parsing duties for a program. In many cases these tools have handle
either approach to parsing as well as validating (see below) a document
as it is read. On IceCube parsing is currently handled by the Xerces-J
tool~\cite{xerces-site} from Apache.

\section{Validation}

An XML document is valid when its contents conform to the rules laid
down in its XML Schema or DTD. As the DTD technology is being superseded
in most cases by the XML Schema alternative it will not be considered
any further in this paper.

Validation should not be confused with a ``well formed'' document. This
simply means that all ``begin"" elements in the document have matching
``end'' elements in the right place in the element hierarchy and that
attributes are correctly delimited. A ``well formed'' document say
nothing about the contents of any elements.

An XML schema describes the allowed contents of a document. It lays down
the rules about how and which elements can be contained inside another
element, as well as defining what attributes an element can
have.

Apart from defining the legal structure of a document, an XML Schema can
also specify which values of element and attributes are considered
legal. Figure~\ref{atwdExample-xsd} is the XML schema against which the
file shown in Figure~\ref{atwdExample-xml} should be validated. The {\tt
enumeration} elements with the {\tt bitsPerSample} attribute are an
example of such a restriction. Full details about XML Schema can be
found elsewhere~\cite{xsd-primer}.

\begin{figure*}[t]
\centering
\begin{verbatim}
<?xml version="1.0" encoding="UTF-8"?>
<xs:schema targetNamespace="http://glacier.lbl.gov/icecube/daq/example"
        xmlns="http://glacier.lbl.gov/icecube/daq/example"
        xmlns:xs="http://www.w3.org/2001/XMLSchema">
  <xs:element name="AtwdReadout">
    <xs:complexType>
      <xs:sequence>
        <xs:element name="Atwd" maxOccurs="unbounded">
          <xs:complexType>
            <xs:sequence>
              <xs:element name="Channel" type="AtwdChannel" maxOccurs="2" minOccurs="2"/>
            </xs:sequence>
          </xs:complexType>
          <xs:unique name="RequireAllChannels">
            <xs:selector xpath="Channel"/>
            <xs:field xpath="@number"/>
          </xs:unique>
        </xs:element>
      </xs:sequence>
    </xs:complexType>
  </xs:element>

  <xs:simpleType name="SixteenBitsList">
    <xs:list itemType="xs:unsignedShort"/>
  </xs:simpleType>

  <xs:simpleType name="AtwdChannelData">
    <xs:restriction base="SixteenBitsList">
      <xs:length value="48"/>
    </xs:restriction>
  </xs:simpleType>

  <xs:complexType name="AtwdChannel">
    <xs:simpleContent>
      <xs:extension base="AtwdChannelData">
        <xs:attribute name="number">
          <xs:simpleType>
            <xs:restriction base="xs:nonNegativeInteger">
              <xs:maxExclusive value="2"/>
            </xs:restriction>
          </xs:simpleType>
        </xs:attribute>
        <xs:attribute default="16" name="bitsPerSample">
          <xs:simpleType>
            <xs:restriction base="xs:nonNegativeInteger">
              <xs:enumeration value="8"/>
              <xs:enumeration value="16"/>
            </xs:restriction>
          </xs:simpleType>
        </xs:attribute>
      </xs:extension>
    </xs:simpleContent>
  </xs:complexType>
</xs:schema>
\end{verbatim}
\caption{The XML schema against which {\tt atwdExample.xml} should be
validated.}
\label{atwdExample-xsd}
\end{figure*}

All this means that it is possible to use XML validation to guarantee
that an input document's content is correct before this content is
passed on to a program for processing. Thus a developer is not required
to write their own validation routines.

\section{Java Data Structures}

As an XML Schema can define data types, Sun has developed the Java
Architecture for XML Binding (JAXB) interface which enables Java classes
to be created by running an XML Schema through a processor.

As well as simple class creation the reference implementation of JAXB
from Sun, which is part of the Java Web Services Developer
Pack~\cite{Java-WSDP}, provides default implementations of these classes
that can easily be read from and written to XML documents.
Figure~\ref{unmarshall-java} is an example of how easy it is to read and
access the contents of the example file shown in
Figure~\ref{atwdExample-xml}.

\begin{figure*}[t]
\centering
\begin{verbatim}
public class Unmarshall {
    // This sample application demonstrates how to unmarshal an instance
    // document into a Java content tree and access data contained within it.
    public static void main( String[] args ) {
        try {

            // create a JAXBContext capable of handling classes generated into
            // the primer.po package
            JAXBContext jc = JAXBContext.newInstance( "icecube.daq" );

            // create an Unmarshaller
            Unmarshaller u = jc.createUnmarshaller();

            // unmarshal a daq instance document into a tree of Java content
            // objects composed of classes from the icecube.daq package.
            AtwdReadout atwdList =
                    (AtwdReadout)u.unmarshal( new FileInputStream( "atwdSample.xml" ) );
            AtwdReadoutType.AtwdType atwd =
                    (AtwdReadoutType.AtwdType)atwdList.getAtwd().get(0);
            System.out.println("Found " + atwd.getChannel().size() +
                    " channels in the first ATWD");
        } catch( JAXBException je ) {
            je.printStackTrace();
        } catch( IOException ioe ) {
            ioe.printStackTrace();
        }
    }
}
\end{verbatim}
\caption{Example code which uses JAXB to read in the {\tt
atwdExample.xml} data file.}
\label{unmarshall-java}
\end{figure*}

\section{Transforms}

One of the powers of XML is that as the structure of a document is
independent of its content it is possible to transform the contents of
one document onto a different document that has an alternate structure.
In IceCube we made use of this feature when we designed the ``Simple
Test Framework'' (STF).

\subsection{Simple Test Framework}

The STF is designed to run a set of modules that test the functionality
of our Digital Optical Module (DOM) hardware. These modules take as
input a set of parameters and provide as a result a set out output
parameters. This whole interface can be described by an XML document.
Moreover the input parameters for a module, when it is to be executed,
can be specified by means of a setup XML file and the output parameters
returned in the form of a results XML file. Figure~\ref{exampleOne-xml}
is an example of a description file, while
Figure~\ref{exampleOneSetup-xml} and~\ref{exampleOneResults-xml} are
examples out setup and results files respectively.

\begin{figure*}[t]
\centering
\begin{verbatim}
<?xml version="1.0" encoding="UTF-8"?>
<stf:test
        xmlns:stf="http://glacier.lbl.gov/icecube/daq/stf"
        xmlns:xsi="http://www.w3.org/2001/XMLSchema-instance"
        xsi:schemaLocation="http://glacier.lbl.gov/icecube/daq/stf stfDefn.xsd">
  <name>ExampleOne</name>
  <description>This is a simple Example of an STF module definition.</description>
  <version major="1" minor="0"/>
  <inputParameter>
    <name>fruit</name>
    <string default="bananas"/>
  </inputParameter>
  <inputParameter>
    <name>quantity</name>
    <unsignedInt default="1" maxValue="100" minValue="0"/>
  </inputParameter>
  <outputParameter>
    <name>fufilled</name>
    <boolean/>
  </outputParameter>
  <outputParameter>
    <name>numberRemaining</name>
    <unsignedInt/>
  </outputParameter>
</stf:test>
\end{verbatim}
\caption{Example description of an STF module.}
\label{exampleOne-xml}
\end{figure*}

\begin{figure*}[t]
\centering
\begin{verbatim}
<?xml version="1.0" encoding="UTF-8"?>
<stf:setup
        xmlns:stf="http://glacier.lbl.gov/icecube/daq/stf"
        xmlns:xsi="http://www.w3.org/2001/XMLSchema-instance"
        xsi:schemaLocation="http://glacier.lbl.gov/icecube/daq/stf stf.xsd">
 <ExampleOne>
   <parameters>
     <fruit>oranges</fruit>
     <quantity>54</quantity>
   </parameters>
 </ExampleOne>
</stf:setup>
\end{verbatim}
\caption{Example setup for running the STF ExampleOne module.}
\label{exampleOneSetup-xml}
\end{figure*}

\begin{figure*}[t]
\centering
\begin{verbatim}
<?xml version="1.0" encoding="UTF-8" ?>
<stf:result
        xmlns:stf="http://glacier.lbl.gov/icecube/daq/stf"
        xmlns:xsi="http://www.w3.org/2001/XMLSchema-instance"
        xsi:schemaLocation="http://glacier.lbl.gov/icecube/daq/stf stf.xsd">
 <ExampleOne>
   <description>
This is a simple Example of an STF module definition.
   </description>
   <version major="1" minor="0"/>
   <parameters>
     <fruit>oranges</fruit>
     <quantity>54</quantity>
     <fufilled>true</fufilled>
     <numberRemaining>19</numberRemaining>
     <passed>true</passed>
     <testRunnable>true</testRunnable>
     <boardID>linux-sim</broadID>
  <ExampleOne>
 <StfEg>
<stf:result>
\end{verbatim}
\caption{Example result from running the STF {\tt ExampleOne} module.}
\label{exampleOneResults-xml}
\end{figure*}

The description file not only associates a type with both input and
output parameters, but can also specify a legal ranges and default value
for input parameters. This sort of information can be used for at least
two different purposes.

\begin{itemize}
\item The creation of a C header file that correctly declares the test
signatures, any appropriate limits, and default values.
\item An XSL Schema file that can be used to validate the input and
output files.
\end{itemize}

In IceCube we have XSL files to handle both of these tasks.
Both of the above tasks can be accomplished by using an ``eXtensible
Stylesheet Language'' (XSL) file and a XSL transform engine, e.g. Xalan
from Apache~\cite{xalan-site}.

\subsection{XSLT}

XSL files are a description of how the contents of an existing XML file
should be transformed by an XSLT to create a new file. This new file is
not required to be XML. It can be a simple text file or a file of
``format objects'' which can be used to generate PDF files.

In IceCube we use XSL files to generate both text files (C header files)
and XML files (XML Schema files). Figure~\ref{exampleOneSignature-h}
contains the C header file that contains the signatures for the {\tt
ExampleOne} module. This file was generated by the XSL in
Figure~\ref{defn2Signature-xsl}.

\begin{figure*}[t]
\centering
\begin{verbatim}
extern BOOLEAN ExampleOneInit(STF_DESCRIPTOR *);
extern BOOLEAN ExampleOneEntry(STF_DESCRIPTOR *,
                   const char* fruit,
                   unsigned int quantity,
                   BOOLEAN*  fufilled,
                   unsigned int*  numberRemaining);
\end{verbatim}
\caption{The resulting signatures generated from {\tt ExampleOne.xml}
using an XSLT.}
\label{exampleOneSignature-h}
\end{figure*}

\begin{figure*}[t]
\centering
\begin{verbatim}
<?xml version="1.0" encoding="UTF-8"?>
<xsl:stylesheet version="1.0" xmlns:stf="http://glacier.lbl.gov/icecube/daq/stf"
        xmlns:xs="http://www.w3.org/2001/XMLSchema"
        xmlns:xsl="http://www.w3.org/1999/XSL/Transform">
  <xsl:output indent="yes" method="text"/>
  <xsl:variable name="nl">
<xsl:text>
</xsl:text>
  </xsl:variable>
  <xsl:template match="/">
    <xsl:apply-templates select="stf:test"/>
  </xsl:template>
  <xsl:template match="stf:test">
  <xsl:variable name="testName" select="name"/>
extern BOOLEAN <xsl:copy-of select="$testName"/>Init(STF_DESCRIPTOR *);
extern BOOLEAN <xsl:copy-of select="$testName"/>Entry(STF_DESCRIPTOR *,
<xsl:apply-templates mode="Entry" select="inputParameter"/>
<xsl:apply-templates mode="Entry" select="outputParameter"/>
  </xsl:template>
  <xsl:template match="stf:test/*/*" mode="Entry">
    <xsl:text>                   </xsl:text>
    <xsl:apply-templates mode="signature" select="."/>
    <xsl:apply-templates mode="entryModifier" select="."/>
    <xsl:text> </xsl:text>
    <xsl:copy-of select="../name"/>
  </xsl:template>
  <xsl:template match="stf:test/*" mode="Entry">
    <xsl:apply-templates mode="Entry" select="boolean|string|unsignedInt|unsignedLong"/>
    <xsl:choose>
      <xsl:when test='((0=count(../outputParameter))or("outputParameter"=local-name()))
              and(last()=position())'>);</xsl:when>
      <xsl:otherwise>,</xsl:otherwise>
    </xsl:choose>
    <xsl:copy-of select="$nl"/>
  </xsl:template>
  <xsl:template match="stf:test/*/boolean" mode="signature">BOOLEAN</xsl:template>
  <xsl:template match="stf:test/inputParameter/string" mode="signature">const char*</xsl:template>
  <xsl:template match="stf:test/outputParameter/string" mode="signature">char*</xsl:template>
  <xsl:template match="stf:test/*/unsignedInt" mode="signature">unsigned int</xsl:template>
  <xsl:template match="stf:test/*/unsignedLong" mode="signature">unsigned long</xsl:template>
  <xsl:template match="stf:test/outputParameter/*" mode="entryModifier">* </xsl:template>
  <xsl:template match="stf:test/*/*" mode="EntryLocal">
    <xsl:text>                   </xsl:text>
    <xsl:apply-templates mode="entryLocalModifier" select="."/>
    <xsl:text>getParamByName(d, "</xsl:text>
    <xsl:copy-of select="../name"/><xsl:text>")->value.</xsl:text>
    <xsl:apply-templates mode="value" select="."/>
    <xsl:text>Value</xsl:text>
  </xsl:template>
  <xsl:template match="stf:test/*" mode="EntryLocal">
    <xsl:apply-templates mode="EntryLocal" select="boolean|string|unsignedInt|unsignedLong"/>
    <xsl:choose>
      <xsl:when test='((0=count(../outputParameter))or("outputParameter"=local-name()))
              and(last()=position())'>);</xsl:when>
      <xsl:otherwise>,</xsl:otherwise>
    </xsl:choose>
    <xsl:copy-of select="$nl"/>
  </xsl:template>
  <xsl:template match="stf:test/outputParameter/*" mode="entryLocalModifier">&amp;</xsl:template>
</xsl:stylesheet>
\end{verbatim}
\caption{The XSL file used to generate a modules signature.}
\label{defn2Signature-xsl}
\end{figure*}

A similar XSL file can be used to create the XML Schema for setup and
result XML files for any given module. This XML Schema can then be used
to verify that the setup file is valid before it is passed to the STF to
be executed, thus relieving the STF of the need to have custom code to
check its input values.

It should also be noted that the XML description has an associated XML
Schema, {\tt stfDefs.xsd} (not show here), against which it should
validate. By requiring any description document to validate against that
schema we can then be confident that both the resultant XML schema file
and C header files will be correct.

\section{Searching}

\subsection{XPath}

The initial form of searching developed for XML documents is XPath. The
basics from of an XPath statement is to specify a hierarchy of elements,
with wildcarding allow. Using this system it is possible to select one
or more elements in a document. This selection can be based on the type
of elements, an elements context or even the value of its attributes or
contents.

XPath is used in XSL documents to identify what transforms should be
applied to which elements. The {\tt match} and {\tt select} attributes
in Figure~\ref{defn2Signature-xsl} are examples of this mechanism.

\subsection{XQuery}

Currently development is underway on an extension of the XPath mechanism
which will span more than one document. This is called XQuery and is
planned to form the basis of a query language for XML databases.

The format of a query in XQuery follows the same lines as XPath and the
resulting elements are collected into a single XML document which acts
as the results of the XQuery.

XML databases are still in their infancy, though there are some examples
in existence~\cite{xindice-site}. At present IceCube is not using one so
no examples of XQuery are currently available.

\section{Data Transfers}

The simple Object Access Protocol (SOAP)~\cite{soap-primer} has become
one of the main engines that has driven the adoption of XML for data
transfers. This system wraps up an XML document in a ``envelope'' with
has a ``header'' and a ``body'' such that the document can be moved from
one process to another. The other process may even be on a different
machine. Non-XML documents can be handled by ``SOAP with
Attachments''~\cite{soap-attachments}.

At this time IceCube has not had the opportunity to pursue the use of
SOAP within its software system, but there are plans to use it as the
main system for handling inter-process communications within the DAQ and
production farms. This will allow components of those sub-systems to be
written in either Java or C++. At present those systems are being
developed in Java and RMI is being used for the inter-process
communications.

\section{Conclusions}

IceCube has been very fortunate in that its circumstances have made it a
good platform in which to assess the promises of the XML technology. XML
is quickly becoming an industry standard for both data and configuration
files. This has mean that there are plenty of tools available to handle
XML. Many of these tools are open source and thus are suitable for use
on IceCube. A lot of work has gone into these tools and their adoption
by IceCube means that we can concentrate on our own issues rather than
the development of tools that are not directly related to our purpose.

Our conclusion is that XML is a good approach for text based files and
small data files. It is easy to understand and customize. There is a
good array of tools available that handle many of the tasks needs to
read and write such files and therefore its adoption can reduce
development time and, because of the tools wide adoption, improve
softwares reliability.

\end{document}